\documentclass[useAMS,usenatbib]{mn2e}
\usepackage{txfonts}

\title[Solar system constraints on $f(T)$ gravity]{Solar system constraints
on $f(T)$ gravity}

\author[Lorenzo Iorio and  Emmanuel N. Saridakis]{Lorenzo
Iorio$^{1}$\thanks{E-mail:
lorenzo.iorio@libero.it; Emmanuel\_Saridakis@baylor.edu} and Emmanuel N.
Saridakis$^{2,3}$\\
$^{1}$I Ministero dell'Istruzione, dell'Universit\`{a} e della
Ricerca (M.I.U.R.), Viale Unit\`{a} di Italia 68
Bari, (BA) 70125,
Italy\\
$^{2}$ Physics Division, National Technical University of Athens,
15780 Zografou Campus,  Athens, Greece\\
$^{3}$ CASPER,
 Physics Department, Baylor University, Waco, TX  76798-7310, USA}

\begin{document}

\date{}


\maketitle

\label{firstpage}

\begin{abstract}
  We use recent observations from solar system orbital motions in order to
constrain $f(T)$ gravity. In particular, imposing a quadratic $f(T)$
correction to the linear-in-$T$ form, which is a good approximation for
every realistic case, we extract the spherical solutions of the theory.
Using these spherical solutions to describe the Sun's gravitational field,
we use recently
determined supplementary advances of planetary perihelia, to infer upper
bounds on the allowed $f(T)$ corrections. We find that the maximal allowed
divergence of the gravitational potential in $f(T)$ gravity from that in
the teleparallel equivalent of General Relativity is of the
order of  $6.2 \times 10^{-10}$, in the applicability region of our
analysis. This is much smaller than the corresponding (significantly small
too) divergence that is predicted from cosmological observations, as
expected.
 Such a tiny allowed divergence from the linear form should be
taken into account in $f(T)$ model building.
\end{abstract}


   \maketitle
%

\section{Introduction}

According to accumulating observations of different kinds, the universe is now known to
be accelerating   (Riess et al.\citep{obs}, Perlmutter et al.
\citep{obs2}). Such a
feature led physicists to follow
two directions in order to explain it. The first one is to introduce
the concept of dark energy (see Copeland  et al. \citep{Copeland:2006wr}
and Bamba  et al. \citep{Bamba:2012cp} and
references therein) in the right-hand-side of the field equations of
General Relativity,
which could either be the simple cosmological constant or various
new exotic ingredients (Ratra    \& Peebles \citep{quint}, Wetterich
\citep{quint1}, Caldwell \citep{phant},   Nojiri    \&
Odintsov    \citep{phant1}, Feng et al.\citep{quintom}, Guo
\citep{quintom1}, Cai et al. \citep{quintom2}). The second
direction is to modify the left-hand-side of the general relativistic field
equations,
that is to modify the gravitational theory itself, with the basic
extended gravitational theories known as $f(R)$-gravity  (see
(De Felice \citep{DeFelice:2010aj}) and references therein). Such
extended theories can present very interesting
behaviors; the corresponding cosmologies have been investigated
in detail (Nojiri  \&
Odintsov  \citep{Nojiri:2006ri}).

A new class of modified-gravity theories, which has
recently received attention by the scientific community, is the $f(T)$
gravity (Ferraro   \& Fiorini \citep{Ferraro:2006jd}, Ferraro   \& Fiorini
\citep{Ferraro:2006jdb}, Bengochea    \&  Ferraro
\citep{Ferraro:2006jdc}, Linder
\citep{Linder:2010py}). It is an extension of
the old
idea of
the ``teleparallel'' equivalent of General Relativity
(TEGR), in which one uses the curvature-less Weitzenb{\"{o}}ck
connection instead of the torsion-less Levi-Civita one, and where
the dynamical objects are the four linearly independent vierbeins
(Einstein \citep{ein28}, Unzicker   \& Case
 \citep{ein28b},  Hayashi     \&  Shirafuji   \citep{Hayashi79}. In
TEGR, the torsion tensor is formed solely from products of first
derivatives of the tetrad. Then, the Lagrangian density $T$ can  be
constructed from such a torsion tensor under the assumptions of invariance
under general coordinate transformations, global Lorentz transformations,
and the parity operation, along with requiring the
Lagrangian density to be second order in the torsion tensor
(Hayashi     \&  Shirafuji \citep{Hayashi79}). Thus, in $f(T)$ gravity one
generalizes the above TEGR
formalism, making the Lagrangian density a function of $T$, similar to the
$f(R)$ extension of Einstein-Hilbert action.

The above $f(T)$-gravitational framework proves to lead to interesting
cosmological behavior and it has gained much attention in the
literature (Myrzakulov \citep{Myrzak1}, Yerzhanov et al
\citep{Myrzak2}, Wu    \&   Yu
\citep{Myrzak3}, Chen et al. \citep{Myrzak4}, Wu \&   Yu
\citep{Myrzak5}, Bamba et al.
\citep{Myrzak6},
Dent et al. \citep{Myrzak7}, Zheng     \& Huang
  \citep{Myrzak8}, Bamba et al. \citep{Myrzak9}, Yang \citep{Myrzak10},
Zhang   et al. \citep{Myrzak11}, Cai et al. \citep{Myrzak12},
Chattopadhyay
   \&  Debnath  \citep{Myrzak13},   Sharif      \& Rani
\citep{Myrzak14},  Wei
 et al. \citep{Myrzak15},   Ferraro   \&   Fiorini
 \citep{Myrzak16}, Wei et al. \citep{Myrzak17}, Capozziello et al.
\citep{Myrzak18}, Wu     \& Yu \citep{Myrzak19},  Bamba      \& Geng
\citep{Myrzak20}, Geng et al. \citep{Myrzak21}, Wei \citep{Myrzak22}, Geng
et
al  \citep{Myrzak23},   Wu    \& Geng
 \citep{Myrzak24}, Bohmer et al. \citep{Myrzak25}, Karami     \&
Abdolmaleki  \citep{Myrzak26},  Atazadeh      \&  Darabi
\citep{Myrzak27},
Farajollahi et al. \citep{Myrzak28},
Yang et al. \citep{Myrzak29},  Karami     \&
Abdolmaleki \citep{Myrzak30},  Karami     \&
Abdolmaleki  \citep{Myrzak31}, Xu et al
\citep{Myrzak32},  Bamba et al. \citep{Myrzak33}, Setare     \&  Houndjo
 \citep{Myrzak34}, Liu et al. \citep{Myrzak35},
  Wu      \&  Yu \citep{Wu:2010mn},
Bengochea \citep{Wu:2010mn1}, Gonzalez et al.
\citep{Gonzalez:2011dr}, Daouda et al.  \citep{Daouda001}, Daouda et al.
\citep{Daouda001b}, Boehmer et al
\citep{Boehmer:2011gw},  Daouda et al. \citep{Daouda:2012nj},   Ferraro
\& Fiorini
\citep{Ferraro001}, Wang
\citep{Wang:2011xf}, Miao et al. \citep{Miao003}, Wei et al
\citep{Wei:2011aa},   Ferraro    \&  Fiorini
\citep{Ferraro:2011ks}). One question that
straightforwardly
arises is what are the qualitative and quantitative $f(T)$-modifications
that are allowed. Although theoretically one has an enhanced freedom, in
practice, in order to confront observations one expects a small divergence
from TEGR, that is from the linear-in-$T$ case which coincides with General
Relativity. Indeed, in Wu      \&  Yu \citep{Wu:2010mn}
and Bengochea \citep{Wu:2010mn1} the authors used data
from cosmological observations in order to constrain the model parameters
of
two well-studied $f(T)$-ansatzes, namely the power-law and the exponential
one, and they found less than 1\% divergence from TEGR.

In the present work we use latest observations of  solar system
orbital motions in order to constrain the parameters of $f(T)$ gravity. In
particular, after extracting the spherical solutions of the theory
we use them to describe the Sun's gravitational field and then we can use
data from planetary motion in order to infer an upper bound on the
allowed divergence from the TEGR.  The solar system analysis imposes
significantly tighter bounds than the cosmological ones
(Wu      \&  Yu \citep{Wu:2010mn},
Bengochea \citep{Wu:2010mn1}) as expected, and in this completely different
observational region it verifies too that the divergences from TEGR (and
thus from General Relativity) are very small.

The plan of the work is as follows: In section \ref{model} we
briefly review $f(T)$ gravity and in section \ref{spersol}
we extract the spherical solutions assuming small   corrections to the
linear-in-$T$ scenario. In section \ref{constrains} we use data form
solar system orbital motions and we impose constraints on the model
parameters. Finally, in section \ref{conclusions} we summarize the
obtained results.

\section{$f(T)$ gravity}
\label{model}

In this section we briefly review $f(T)$ gravity. Our
notation is as
follows: Greek indices $\mu, \nu,$...
and capital Latin indices $A, B, $...
run over all coordinate and tangent
space-time 0, 1, 2, 3, while lower case Latin indices (from the middle of
the
alphabet) $i, j,...$ and lower case Latin indices
(from the beginning of the alphabet) $a,b, $...
run over spatial and tangent
space coordinates 1, 2, 3, respectively.

As stated in the Introduction, the dynamical variable of
``teleparallel'' gravity, as well as of its $f(T)$ extension, is the
vierbein
field ${\mathbf{e}_A(x^\mu)}$. This forms an orthonormal basis
for the tangent
space at each point $x^\mu$ of the manifold, that is $\mathbf{e}
_A\cdot%
\mathbf{e}_B=\eta_{AB}$, where $\eta_{AB}=diag (1,-1,-1,-1)$.
Furthermore,
the vector $\mathbf{e}_A$ can be analyzed with the use of its
components $%
e_A^\mu$ in a coordinate basis, that is
$\mathbf{e}_A=e^\mu_A\partial_\mu $.

In such a construction, the metric tensor is obtained from the
dual vierbein
as
\begin{equation}  \label{metrdef}
g_{\mu\nu}(x)=\eta_{AB}\, e^A_\mu (x)\, e^B_\nu (x).
\end{equation}
Contrary to General Relativity, which uses the torsion-less
Levi-Civita
connection, in the present formalism ones uses the curvature-less
Weitzenb%
\"{o}ck connection
${\Gamma}^{\lambda\{w\}}_{\nu\mu}\equiv e^\lambda_A\:
\partial_\mu
e^A_\nu$  (Weitzenb\"{o}ck \citep{Weitzenb23}), whose torsion tensor reads
\begin{equation}  \label{torsion2}
{T}^\lambda_{\:\mu\nu}={\Gamma}^{\lambda\{w\}}_{
\nu\mu}-%
{\Gamma}^{\lambda\{w\}}_{\mu\nu}
=e^\lambda_A\:(\partial_\mu
e^A_\nu-\partial_\nu e^A_\mu).
\end{equation}
Moreover, the contorsion tensor, which equals to the difference
between Weitzenb%
\"{o}ck and Levi-Civita connections, is defined as
$K^{\mu\nu}_{\:\:\:\:\rho}\equiv-\frac{1}{2}\Big(T^{\mu\nu}_{
\:\:\:\:\rho}
-T^{\nu\mu}_{\:\:\:\:\rho}-T_{\rho}^{\:\:\:\:\mu\nu}\Big)$  and we also
define
$
S_\rho^{\:\:\:\mu\nu}\equiv\frac{1}{2}\Big(K^{\mu\nu}_{\:\:\:\:\rho}
+\delta^\mu_\rho
\:T^{\alpha\nu}_{\:\:\:\:\alpha}-\delta^\nu_\rho\:
T^{\alpha\mu}_{\:\:\:\:\alpha}\Big)$.

Using these quantities one can define the teleparallel Lagrangian, which is
the torsion scalar, as
(Hayashi     \&
Shirafuji \citep{Hayashi79}, Maluf \citep{Maluf:1994ji},
 Arcos     \&  Pereira  \citep{Maluf:1994jib})
\begin{equation}  \label{telelag}
T\equiv S_\rho^{\:\:\:\mu\nu}\:T^\rho_{\:\:\:\mu\nu}.
\end{equation}
In summary, in the present formalism all the information
concerning the
gravitational field is included in the torsion tensor ${T}%
^\lambda_{\:\mu\nu} $, and the torsion scalar $T$ arises from it
in a
similar way as the curvature scalar arises from the curvature
(Riemann)
tensor.

While in the teleparallel equivalent of General Relativity (TEGR) the
action is just $T$, the idea of $f(T)$ gravity is to generalize $T$
to a
function $T+f(T)$, which is similar in spirit to the
generalization of the
Ricci scalar $R$ in the Einstein-Hilbert action to a function
$f(R)$. In
particular, the action in a universe governed by $f(T)$ gravity
reads:
\begin{eqnarray}  \label{action}
I = \frac{1}{16\pi G}\int d^4x e \left[T+f(T)-2\Lambda+L_m\right],
\end{eqnarray}
where $e = \det(e_{\mu}^A) = \sqrt{-g}$,
$G$ is the Newton's
constant (we also set for convenience the light speed to one),  $L_m$
stands
for the
matter Lagrangian, and we have added for completeness a cosmological
constant $\Lambda$ (which could alternatively be absorbed into $f(T)$). We
mention here that since the Ricci scalar
$R$ and the torsion scalar $T$
differ only by a total derivative  (Weinberg \citep{Weinberg:2008}), in the
case where $f(T)$ is zero
the action (\ref{action}) is equivalent to General Relativity
with a cosmological constant.

Variation of the action
(\ref{action}) with
respect to the vierbein gives the equations of motion
\begin{eqnarray}\label{eom}
e^{-1}\partial_{\mu}(ee_A^{\rho}S_{\rho}{}^{\mu\nu})[1+f_{,T}]
-e_{A}^{\lambda}T^{\rho}{}_{\mu\lambda}S_{\rho}{}^{\nu\mu} +
e_A^{\rho}S_{\rho}{}^{\mu\nu}\partial_{\mu}({T})f_{,TT}\nonumber\\
-\frac{1}
{ 4 } e_ { A } ^ {
\nu
}[T+f({T})-2\Lambda]
= 4\pi Ge_{A}^{\rho}T_{\rho}{}^{\nu\{em\}},
\end{eqnarray}
where $f_{,T}$ and $f_{,TT}$  denote respectively  the
first and second derivatives of the function $f(T)$ with respect
to $T$. Note that the tensor
${T%
}_{\rho}{}^{\nu\{em\}}$ on the right-hand side is the usual
energy-momentum tensor.

\section{Spherically symmetric Solutions}
\label{spersol}

We are interesting in extracting spherically symmetric solutions of the
four-dimensional $f(T)$ gravity presented in the previous section, and in
particular to extract the correction to the Schwarzschild solution of
General Relativity. Up to now in the literature it has been done
in three dimensions  (Gonzalez et al. \citep{Gonzalez:2011dr}) and
partially in four dimensions
for specific $f(T)$ ansatzes
 (Daouda et al.  \citep{Daouda001}, Daouda et al.
\citep{Daouda001b}, Boehmer et al.
\citep{Boehmer:2011gw},  Daouda et al. \citep{Daouda:2012nj}), while there
is also a different approach, namely to find the general (non-diagonal)
vierbein choice that corresponds exactly to the Schwarzschild solution
(Ferraro    \&  Fiorini
\citep{Ferraro:2011ks}, Ferraro
\& Fiorini
\citep{Ferraro001},  Wang
\citep{Wang:2011xf},  Maluf et al. \citep{Maluf:2012na}).

Let us consider as usual the general metric ansatz of the form
\begin{eqnarray}
  ds^2 = N(r)^2dt^2-K(r)^{-2}dr^2 -R(r)^2 d\Omega^2\,,
  \label{sphermetric}
\end{eqnarray}
where $d\Omega^2 = d\theta^2 + \sin^2  \theta\, d\varphi^2$ and
where $N(r)$, $K(r)$ and $R(r)$ are three unknown functions. Note that we
do not restrict in the case $N(r)=K(r)$ as in (Daouda et al.
\citep{Daouda:2012nj}), since
we desire to be as general as possible. As a next step we
need to extract the corresponding vierbeins that give rise to
the above metric, through relation (\ref{metrdef}). The simplest choice is
to use
a diagonal vierbein of the form
\begin{equation}
  e^A_{\mu} =
diag
\left(N(r),K(r)^{-1},R(r),R(r)\sin \theta\right).
  \label{vierb1}
\end{equation}
 However, we mention that although in the case of linear-in-$T$
gravity, such a simple relation between the metric and the vierbeins is
always allowed, in the general $f(T)$ gravity  this is not
the case anymore, and in general one has a more complicated relation
connecting the vierbein tetrad with the metric, with the former being
non-diagonal even for a diagonal metric (Sotiriou et al.
\citep{fTLorinv0}),
as it arises
from Lorentz transformations of (\ref{vierb1}). Nevertheless, in the
4D cosmological investigations of $f(T)$ gravity
(Ferraro   \& Fiorini \citep{Ferraro:2006jd}, Ferraro   \& Fiorini
\citep{Ferraro:2006jdb}, Bengochea    \&  Ferraro
\citep{Ferraro:2006jdc}, Linder
\citep{Linder:2010py}, Myrzakulov \citep{Myrzak1}, Yerzhanov et al
\citep{Myrzak2}, Wu    \&   Yu
\citep{Myrzak3}, Chen et al. \citep{Myrzak4}, Wu \&   Yu
\citep{Myrzak5}, Bamba et al.
\citep{Myrzak6},
Dent et al. \citep{Myrzak7}, Zheng     \& Huang
  \citep{Myrzak8}, Bamba et al. \citep{Myrzak9}, Yang \citep{Myrzak10},
Zhang   et al. \citep{Myrzak11}, Cai et al. \citep{Myrzak12},
Chattopadhyay
   \&  Debnath  \citep{Myrzak13},   Sharif      \& Rani
\citep{Myrzak14},  Wei
 et al. \citep{Myrzak15},   Ferraro   \&   Fiorini
 \citep{Myrzak16}, Wei et al. \citep{Myrzak17}, Capozziello et al.
\citep{Myrzak18}, Wu     \& Yu \citep{Myrzak19},  Bamba      \& Geng
\citep{Myrzak20}, Geng et al. \citep{Myrzak21}, Wei \citep{Myrzak22}, Geng
et
al  \citep{Myrzak23},   Wu    \& Geng
 \citep{Myrzak24}, Bohmer et al. \citep{Myrzak25}, Karami     \&
Abdolmaleki  \citep{Myrzak26},  Atazadeh      \&  Darabi
\citep{Myrzak27},
Farajollahi et al. \citep{Myrzak28},
Yang et al. \citep{Myrzak29},  Karami     \&
Abdolmaleki \citep{Myrzak30},  Karami     \&
Abdolmaleki  \citep{Myrzak31}, Xu et al
\citep{Myrzak32},  Bamba et al. \citep{Myrzak33}, Setare     \&  Houndjo
 \citep{Myrzak34}, Liu et al. \citep{Myrzak35},
 Wu      \&  Yu \citep{Wu:2010mn},
Bengochea \citep{Wu:2010mn1},
 Ferraro
\& Fiorini
\citep{Ferraro001},
 Daouda et al.  \citep{Daouda001}, Daouda et
al.
\citep{Daouda001b}), as well as in its
black
hole solutions
(Wang
\citep{Wang:2011xf},    Miao et al. \citep{Miao003}, Wei et al
\citep{Wei:2011aa},   Ferraro    \&  Fiorini
\citep{Ferraro:2011ks}), the
authors still use the simple relation between the vierbeins and the
metric, as a first approach to reveal the structure and the features of the
theory. Therefore, in the present work, we also
assume the simple relation between the vierbeins and the metric as a first
approach on this novel theory, capable of revealing the main features
of the solutions. Clearly a detailed investigation of the general vierbein
choice in   $f(T)$ gravity, and its relation to extra degrees of
freedom, is a necessary step for the understanding of this new theory
(Cai  et al. \citep{inprep}).

Inserting the vierbein choice (\ref{vierb1})  into the torsion scalar
definition
(\ref{telelag}) we obtain
\begin{equation}
  T(r) = 2K^2\frac{R'}{R} \left(\frac{2N'}{N}+\frac{R'}{R}\right)
  \label{Torsionsca},
\end{equation}
where primes denote derivatives with respect to $r$. Thus, substitution in
the  field equations (\ref{eom}), neglecting the matter energy-momentum
tensor, provides the following separate equations of motion:
\begin{eqnarray}
&&0=K^2\frac{R'}{R}T'
f_{TT}-\left(\frac{T+f-2\Lambda}{4}\right)\nonumber\\
&&\  \ \ \ \ \ +\left\{T-\frac{
1}{R^2}-K^2\left[\left(\frac{2N'}{N}-\frac{2K'}{K}\right)\frac{R'}{R}
-2\frac { R'' } { R } \right] \right\}
\left(\frac{
1+f_T}{2}\right),
  \label{eq1}\\
 &&0= \left(\frac{1}{R^2}-T\right)\left(\frac{
1+f_T}{2}\right)+\left(\frac{T+f-2\Lambda}{4}\right),
  \label{eq2}\\
 &&0=
-K^2\left(\frac{N'}{N}+\frac{R'}{R}\right)T'f_{TT}+\left(\frac{T+f-2\Lambda
}{4}\right)\nonumber\\
 & &\ \ \ \ \  \ \ \ \  \ \ \   \ \ \ \ -\left\{ \frac{T}{2}
+K^2\left[\frac{R''}{
R}+\frac{N''}{N}-\frac{N'^2}{N^2}\right.\right.\nonumber\\
&& \ \ \ \ \  \ \ \ \  \ \ \ \  \ \ \ \ \ \ \ \ \  \ \ \ \  \ \ \ \
\left.\left.+\left(\frac{N'}{ N } +\frac { R' } {
R}\right)\left(\frac{N'}{N}+\frac{K'}{K}\right)\right]\right\}  \left(\frac
{
1+f_T}{2}\right).
  \label{eq3}
\end{eqnarray}
Finally, we mention here that while in teleparallel gravity, as well as in
General Relativity, the off-diagonal components of the field equations
vanish, in $f(T)$ gravity even the diagonal vierbein field
gives rise to an extra equation (Boehmer et al. \citep{Boehmer:2011gw}). In
particular, the
$r$-$\theta$ equation reads:
\begin{eqnarray}
  \frac{K\cot\theta}{2R^2}T'f_{TT}=0\,.
  \label{eq4}
\end{eqnarray}

The above equations  (\ref{Torsionsca})-(\ref{eq4}) cannot obtain
analytical solutions in general. Thus,
we proceed making two assumptions. The fist is the usual one in order to
reduce the unknown functions, namely without loss of generality we set
$R(r)=r$ (going to more general $R(r)$ will only lead to more complicated
expressions but qualitatively the results will be similar). The second
assumption is related to the $f(T)$ ansatz. In particular, since the usual
General Relativity is obtained by setting $f(T)$ to zero, it is expected
that a realistic $f(T)$ must be small comparing to $T$, and as we mentioned
this has been verified using cosmological observations
(Wu      \&  Yu \citep{Wu:2010mn},
Bengochea \citep{Wu:2010mn1}).
Therefore, by expanding every
$f(T)$ ansatz in $T$-powers, we deduce that as a first approximation we can
consider a
form of
\begin{eqnarray}
\label{fTansatz}
f(T)=\alpha T^2+{\cal{O}}(T^3),
\end{eqnarray}
where we do not write the constant and the linear terms in
the above expansion since they can  be respectively absorbed in $\Lambda$
and $T$ terms of the action (\ref{action}). In the above expression
$\alpha$ is a pure $f(T)$-parameter that determines at first
approximation the divergence from teleparallel gravity, that is from
General Relativity. We mention here that in
Friedmann-Robertson-Walker metric $T=-6H^2$ (Linder \citep{Linder:2010py}),
while in
the Schwarzschild metric $T\propto\Lambda$  for large $r$
(Aldrovandi  \&  Pereira  \citep{Aldrovandi},   Ferraro    \&  Fiorini
 \citep{Ferraro:2011ks}), that is in
both cases $T\ll1$ in S.I units, which is an additional indication that the
above expansion is justified (moreover that is why we do not consider
ansatzes of negative power laws of $T$).
Finally, the above ansatz has been used as a first non-linear correction of
$f(T)$ gravity in
 (Ferraro    \&  Fiorini \citep{Ferraro:2011ks}, Gonzalez et al.
\citep{Gonzalez:2011dr}, Daouda et al. \citep{Daouda:2012nj}). In summary,
going beyond the assumption (\ref{fTansatz}), although it would bring
significant mathematical complications, it would not qualitatively change
our results.

Thus, under the ansatz (\ref{fTansatz}),   equations
(\ref{eq1})-(\ref{eq4}) up to ${\cal{O}}\left(\alpha/r^2\right)^2$
lead to the solution:
\begin{eqnarray}
\label{sol}
&&N(r)^2=1-\frac{2GM}{c^2r}-\frac{\Lambda}{3}r^2\ \ \ \ \ \ \ \ \ \ \
\ \ \ \ \ \ \ \ \ \ \ \ \ \nonumber\\
&&\ \ \ \ \ \ \ \ \ \ \ \ +\alpha\left[
-6\Lambda-\frac{6}{r^2}-\frac{4GM\Lambda}{c^2r}\right]\\
&&K(r)^2=1-\frac{2GM}{c^2r}-\frac{\Lambda}{3}r^2 \ \ \ \ \ \ \ \ \ \ \
\ \ \ \ \ \ \ \ \ \ \ \ \  \ \ \ \ \ \ \ \ \ \ \
 \ \  \nonumber\\
&&\ \ \ \ \ \ +\alpha\left[
 \frac{8\Lambda}{3}-\frac{14}{r^2}-2\Lambda^2r^2-\frac{2GM}{c^2r}
\left(8\Lambda-\frac{8}{r^2}\right)\right],\ \ \label{solK}
\end{eqnarray}
where we have restored the light speed, and where we have introduced
the
integration constant $GM$, with    $M$
the mass of the spherical object. We mention here that in the above
solutions we have kept only up to linear terms in $\alpha/r^2$, since
higher-order terms are expected to be extremely small for $r$ in the
 region of the solar system that we have orbital data. In any case, this
assumption will be checked  a posteriori, after we extract the bounds
on $\alpha$. Finally,  we straightforwardly observe
that in the limit $\alpha\rightarrow0$, the above solution coincides with
the Schwarzschild-de Sitter one (Rindler \citep{Rindler:2006km}) as
expected, while
for $\alpha,\Lambda\rightarrow0$ it becomes the usual Schwarzschild one.

\section{Solar system constraints}
\label{constrains}

In the previous section we extracted the spherically symmetric solutions of
$f(T)$
gravity, in the case where there is a small but general divergence from
the $T$-gravity, that is from the teleparallel equivalent of General
Relativity (TEGR). Therefore, in the present section we apply these
solutions to the gravitational field of the Sun, and we use the latest
solar system data from planetary orbital motions (Fienga \citep{fienga011})
to perform a sensitivity analysis on the two model parameters, namely the
cosmological constant $\Lambda$ and the parameter $\alpha$ that determines
the divergence from General Relativity. Finally, as usual we neglect the
effects of spatial curvature, which is a robust approximation for solar
system scales.

As usual, for a gravitational field determined by a spherically symmetric
metric $g_{\mu\nu}$, the  Newtonian potential and its corrections are
extracted from the $00$ component, namely if $g_{00}=1+h_{00}$, where
$h_{00}$
is dimensionless, then the  potential is
$U = h_{00} c^2/2$. Therefore, from the $g_{00}$ form of $f(T)$ gravity
spherical solutions, that is from $N(r)^2$ of (\ref{sol}), we observe that
there are two corrections to the Newtonian potential $U_{\rm N}$ affecting
a test
particle's orbit, namely
\begin{eqnarray}
&&U_{\Lambda} \label{ul}  = -\frac{\Lambda c^2 r^2}{6} \\
&&U_{\alpha} \label{ua}  = -\frac{3\alpha c^2}{r^2}.
\end{eqnarray}
We mention here that the term  $-6\alpha\Lambda$ is uniform, that is it
does not depend on $r$, and therefore it does not affect orbital
motions, while the term $\propto\alpha\Lambda GM/r$ does not yield secular
orbital precessions, since it just corresponds  to a Newtonian term with
$GM$
rescaled. Thus, in the following it is adequate to consider only the
terms (\ref{ul}),(\ref{ua}). Lastly, note that the
dimensions of the two parameters are
\begin{eqnarray}
&&[\Lambda]  =  {\rm L}^{-2},   \\
&&[\alpha] = {\rm L}^2.
\end{eqnarray}

The orbital effects of extra-potentials having the functional forms of
(\ref{ul}),(\ref{ua}) have been incorporated analytically several times
with a variety of different approaches in the framework of solar system
investigations
(Islam
\citep{Islam83}, Cardona    \&
Tejeiro \citep{Islam83b}, Kerr et al. \citep{Islam83c},
 Kraniotis     \&   Whitehouse  \citep{Islam83d},  Jetzer     \& Sereno
\citep{Islam83e}, Kagramanova et al. \citep{Islam83f}, Sereno     \&
Jetzer  \citep{Islam83g}, Adkins et al.
\citep{Islam83h},
Adkins et al. \citep{Islam83i},   Sereno    \&   Jetzer
 \citep{Islam83j},  Iorio    \citep{Iorio06}, Iorio  \citep{ Islam83l},
Iorio
 \citep{Islam83m}) , and they can be
straightforwardly handled using, for instance, the standard Lagrange
perturbative scheme
(Bertotti et al. \citep{BeFa}). From the equations for the variations of
the
osculating
Keplerian orbital elements (Bertotti et al. \citep{BeFa}) we immediately
deduce that
only the longitude of the pericenter $\varpi\doteq
\Omega + \omega$
 (with $\Omega$ the longitude of the
ascending node  and $\omega$  the argument of pericenter) and the mean
anomaly $\mathcal{M}$ undergo secular
precessions, due to the spherical symmetry of (\ref{ul}),(\ref{ua}) and
their time-independence. On the other hand, the   semimajor axis $a$, the
eccentricity $e$, the inclination $I$ of the orbital plane to the reference
$\{x,y\}$ plane chosen, and the longitude of the ascending node $\Omega$,
remain unaffected
\footnote{Indeed, the Lagrange rate equation for $a$
(Bertotti et al. \citep{BeFa}) contains the partial derivative of the
averaged perturbing
potential $\left\langle U_{\rm pert}\right\rangle$ with respect to
$\mathcal{M}$, which is
proportional to $t$. The Lagrange equations for $e,I,\Omega$ (Bertotti et
al. \citep{BeFa})
are formed with the partial derivatives of $\left\langle U_{\rm pert}\right\rangle$ with
respect to $\Omega,\omega, I$ (and $\mathcal{M}$ as well in the Lagrange
equation for $e$), which are absent in spherically symmetric perturbing
potentials.}. However, from the point of view of a comparison with
solar system
observations, we are interested only in the secular precessions of the
perihelia, since
the rates of the mean anomaly are dominated by the further, additive
mismodeling in the Keplerian mean motions $n_{\rm b}\doteq\sqrt{GM a^{-3}}$
due to the uncertainty  $\sigma_{GM}=10\ {\rm km^3\
s^{-2}}$ in the solar gravitational parameter $GM$
(Konopliv et al. \citep{konopliv011}).

The Lagrange equation for the secular precession of the longitude of
pericenter is (Bertotti et al. \citep{BeFa}):
{\begin{equation}
\left\langle\frac{d\varpi}{d t}\right\rangle = -\frac{1}{n_{
b}
a^2}\left\{\left[\frac{\left(1-e^2\right)^{1/2}}{e}\right]\frac{
\partial\left\langle\mathcal{R}\right\rangle}{\partial e} +
\frac{\tan\left(I/2\right)}{\left(1-e^2\right)^{1/2}}\frac{
\partial
\left\langle\mathcal{R}\right\rangle}{\partial I}\right\}, \label{lagra}
 \end{equation}
where
{\begin{equation}
\mathcal{R}=\left\langle U_{\rm pert}\right\rangle \end{equation}
is the average of the perturbing potential over one  full orbital
revolution of the test particle.
Inserting  (\ref{ul}),(\ref{ua}) into  (\ref{lagra})
we obtain
\begin{eqnarray}
&&\dot\varpi_{\Lambda} \label{periL}
  = \frac{1}{2}\left(\frac{\Lambda
c^2}{n_{\rm
b}}\right)\sqrt{1-e^2}, \\ \nonumber \\
&&\dot\varpi_{\alpha} \label{peria}  = \frac{3\alpha c^2}{a^4 n_{\rm
b}\left(1-e^2\right)},
\end{eqnarray}
where from now on for conciseness we neglect the
brackets $\left\langle\ldots\right\rangle$ denoting the average over one
orbital period, while the overdots indicate the time derivative.

\begin{table}
\centering
\bigskip
\caption{Supplementary rates of change $\Delta\dot\Omega$  and
$\Delta\dot\varpi$, of nodes $\Omega$
 and perihelia $\varpi$ respectively, of
some planets of the solar system, in milliarcseconds
per century (mas cty$^{-1}$), estimated by Fienga
et al. \citep{fienga011} with the INPOP10a ephemerides. Data from Messenger
and Cassini were included for Mercury and Saturn. The reference $\{x,y\}$
plane is the mean Earth's equator at J$2000.0$.
}\label{tavoletta}
\begin{tabular}{lll}
\hline\noalign{\smallskip}
&   $\Delta\dot \Omega$ $\left(\frac{\rm mas}{\rm cty}\right)$ &
$\Delta\dot \varpi $
$\left(\frac{\rm
mas}{\rm cty}\right)$  \\
\noalign{\smallskip}\hline\noalign{\smallskip}
Mercury & $1.4 \pm 1.8$ & $0.4 \pm 0.6$ \\
Venus & $0.2 \pm 1.5$ & $ 0.2\pm 1.5$ \\
Earth & $0.0\pm 0.9$ & $-0.2\pm 0.9$ \\
Mars & $-0.05\pm 0.13$ & $-0.04\pm 0.15$ \\
Jupiter & $-40\pm 42$ & $-41\pm 42$ \\
Saturn & $-0.1\pm 0.4$ & $0.15\pm 0.65$ \\
\noalign{\smallskip}\hline\noalign{\smallskip}
\end{tabular}
\end{table}
In order to preliminarily\footnote{\label{fot}An alternative approach would
consist of explicitly modeling  $f(T)$ gravity in the planetary data
processing softwares of the ephemerides, and fitting such ad-hoc modified
dynamical models to the same observations. Thus, $\alpha$ and $\Lambda$
would be estimated as solve-for parameters.} infer upper bounds on both
$\Lambda$ and
$\alpha$, we compare  (\ref{periL}),(\ref{peria}) for some
planets
to the supplementary rates of change  $\Delta\dot\varpi$  of their
perihelia  $\varpi$, listed in Table \ref{tavoletta}, determined
from observations by Fienga et al.
\citep{fienga011} with the
INPO10a ephemerides.

However, one must be careful in straightforwardly comparing
(\ref{periL}),(\ref{peria}) to the extra-rates of Table \ref{tavoletta},
since $\Delta\dot\varpi$, by construction, account for all those competing
forces which were not modeled\footnote{We only deal  with
well known and established forces, both Newtonian and Einsteinian, since it
would
be pointless to invoke the potentially disturbing action of any sort of
putative, exotic effect, whose alleged biasing
action may,
after all, be present even if a full covariance analysis is performed, as
outlined in footnote \ref{fot}.}
in the software of the INPOP10a
ephemerides.
Moreover, $\Delta\dot\varpi$ are also affected  by the
unavoidable mismodeling in the effects which were actually included in the
models fitted to the observations in INPOP10a. Thus, one should correct
$\Delta\dot\varpi$ for those unmodeled/mismodeled dynamical features of
motion, whose
magnitude is at least comparable to the uncertainties in determining
$\Delta\dot\varpi$ themselves. In this respect, the actions of the Sun's
quadrupole mass moment $J_2$ (Rozelot    \& Damiani \citep{Roz011}),
currently known with a $10\%$
uncertainty, and the general relativistic Lense-Thirring effect
(Lense   \&  Thirring \citep{LT18}), caused by the Sun's angular momentum
$S$ and not modeled at
all in INPOP10a, are relevant since they induce competing secular
perihelion precessions whose size is large enough, as far as the inner
planets, whose extra-rates are more accurately determined, are concerned.
The explicit analytical expressions, valid for a generic spatial
orientation of the Sun's rotational axis (actually it does not point
exactly to the North Celestial Pole), of the perihelion precessions due
to $J_2$ and the Lense-Thirring effect, can be found in
Iorio \citep{IorioPRD}.

Having at our disposal the extra-precessions of the perihelia of more than
one planet, it is possible to set up linear combinations of the form
\begin{eqnarray}
\label{equaz}
 &&\dot\varpi_{\rm A} = \Lambda k_{\rm A}^{(\Lambda)} +
\alpha k_{\rm
A}^{(\alpha)}+ J_2 k_{\rm A}^{(J_2)}+ S k_{\rm A}^{(\rm LT)},\\
&&\ \, \ {\rm
A=Mercury,\ Venus,\ldots},
\nonumber
 \end{eqnarray}
where the coefficients $k^{(\ldots)}_{\rm A}$ come from the analytical
expressions of the precessions due to the various effects considered
similar to (\ref{periL}),(\ref{peria}).
In summary, it is possible to construct a  non-homogeneous linear
system of four equations for the four unknowns $\Lambda,\alpha,J_2,S$,
which is solvable. The resulting values of $\Lambda$ and $\alpha$ are,
by construction,
independent of both $J_2$ and  $S$, whatever their values may be.
By propagating the uncertainties in $\Delta\dot\varpi_{\rm A}$ entering the
expressions of $\Lambda$ and $\alpha$, it is possible to preliminarily
constrain them.
By using the first four inner planets, it turns out that
\begin{eqnarray}
|\Lambda| \label{upL} & \leq 6.1\times 10^{-42}\ {\rm m^{-2}}, \\ \nonumber
\\
|\alpha| \label{upa} & \leq 1.8\times 10^4  \ {\rm m^2}.
\end{eqnarray}
It may be noticed that the bound on $\Lambda$ of (\ref{upL}) is tighter
than the one in Iorio \citep{Iorio06} by  about one order of magnitude.

Since we have now extracted the upper bound on the  $f(T)$-correction
allowed from solar system orbital motions, we can examine the accuracy of
the approximations made in our analysis. Firstly, taking $r$ to be the
Mercury mean heliocentric distance
$\overline{r}_{\rm Merc} = a\left(1 +
\frac{e^2}{2}\right)=5.9\times 10^{10}\ {\rm m}$ ($a$ is its semimajor
axis and $e$ its
  eccentricity), which is the
smallest distance where we have observational data, we obtain
$\alpha/{\overline{r}^2_{\rm Merc}}\approx 5\times 10^{-18}$, and this
justifies our
approximation to neglect orders of $\mathcal{O}\left(\alpha/r^2\right)^2$ in
 (\ref{sol}),(\ref{solK}). Additionally, given the small value of the
cosmological constant, we indeed see that the two cross-terms of
(\ref{sol}), which are proportional to $\alpha\Lambda$, that we did not
take into account in the analysis, are many orders of magnitude smaller
that the terms  (\ref{ul}),(\ref{ua}) that we did use (whether we use
Mercury's data, that is enhancing the $\alpha$-term, or the Pluto's data,
that is enhancing the $\Lambda$-term), and thus our approximation is
well-justified. Moreover, calculating $T$ from (\ref{Torsionsca}) we find
that $T\approx-2\Lambda(1+6\alpha\Lambda)$ for $r\gg1$ (in S.I units), a
result that
verifies our indication that $T\ll1$ (in S.I units) in (\ref{fTansatz}).
Finally, we can
also verify that $\alpha T^2\ll T$, that is the correction to the linear
term is very small as expected.

Relation (\ref{upa}) is the main result of the present work, that is it is
the upper bound of the $f(T)$-correction to teleparallel
gravity (that is to General Relativity), allowed from solar system orbital
motions. In order to provide this allowed correction in a more transparent
way, we denote by $ \Delta U_{f(T)}$ the  divergence of the
gravitational potential in $f(T)$ gravity from that in the teleparallel
equivalent of General Relativity, defined as the ratio of the correction
term $6|\alpha| r^{-2}$ in (\ref{sol}) to the standard term $2GM c^{-2}
r^{-1}$. The maximal allowed value of  $ \Delta U_{f(T)}$ will be acquired 
at the Mercury's orbit $r=\overline{r}_{\rm Merc}$, in
which the correction term is the largest one, since $\overline{r}_{\rm
Merc}$ is the smallest distance in the region where data
exist. Thus, we find
\begin{eqnarray}
\label{fTcor}
 \Delta
U_{f(T)}\Big|_{r=\overline{r}_{\rm Merc}}\approx
\frac{6|\alpha|
\overline{r}_{\rm Merc}^{-2}}{2GM c^{-2}\,
\overline{r}_{\rm Merc}^{-1}}\lesssim 6.2 \times 10^{-10};
 \end{eqnarray}
as expected the allowed $f(T)$-correction affecting solar system
orbital motions is very small.

 \section{Conclusions}
\label{conclusions}

In this work we used latest data from solar system orbital
motions to constrain $f(T)$ gravity. In particular, considering
the basic and usual ansatz $f(T)=\alpha T^2$, which is a good
approximation in all realistic cases, and including also a cosmological
constant $\Lambda$ for completeness, we extracted the spherically symmetric
solutions
of the theory, which coincide with the Schwarzschild-de Sitter one in the
limit $\alpha\rightarrow0$. Thus, by describing the Sun's exterior
gravitational field
by these solutions, we were able to use data from planetary motions in
order to constrain $\alpha$ and $\Lambda$.

Concerning the cosmological constant $\Lambda$, we obtained the usual tiny
bounds. Interestingly enough, our current $f(T)$-analysis leads to one
order of magnitude tighter constraints than General Relativity
\citep{Iorio06}, that is without considering the $\alpha$-term in the
metric. Concerning the pure $f(T)$-parameter $\alpha$, the
obtained bound  (\ref{upa}) leads to a maximal allowed divergence   of
the gravitational potential in $f(T)$ gravity from that in the
teleparallel equivalent of General
Relativity of the order of  $
\lesssim 6.2 \times 10^{-10}$ for the
smallest distance $\overline{r}_{\rm Merc}$ (Mercury mean heliocentric
distance) that we have data, and where the
correction term is the largest one.

The above obtained small divergence from General Relativity is much smaller
than the corresponding (significantly small too) divergence that is
predicted from cosmological constrains  (Wu      \&  Yu \citep{Wu:2010mn},
Bengochea \citep{Wu:2010mn1}), where one also finds that $f(T)$ must be
close to the linear-in-$T$ form. This is standard in observational
constraints and justified, since the solar system observations are always
more accurate than the cosmological ones.   In summary,
in the present analysis we did verify, in a different context, the expected
result that the allowed
divergences of $f(T)$ gravity from the linear (teleparallel) form are
significantly small, and this should be taken into account in $f(T)$ model
building.

In the above analysis we remained in the diagonal vierbein ansatz, since
in this case one can safely elaborate the spherical solutions. However,
observing the features of some specific spherical solutions under  
non-diagonal ansatzes (Daouda et al. \citep{Daouda:2012nj}), in which one
obtains similar terms with the present expressions, we deduce that
the above analysis would lead to qualitatively similar results even in
those cases. However, since for the moment it is not clear how to elaborate
the non-diagonal cases and in particular how to handle the extra degrees of
freedom at the background and especially at the perturbation level, we
preferred  to remain in the diagonal vierbein scenario.

We close this work by mentioning that since in the solutions of the
aforementioned analysis there appear terms of the
form $\alpha/r^2$ (which are small for planetary motions as discussed
above and even a full Parametrized-Post-Newtonian (PPN) analysis
(Will \citep{Will:2005va}) will not change the obtained results), they
could be significant at much smaller distances, outside the
applicability region of the present analysis. Thus, an interesting and
necessary investigation would be to constrain the allowed correction of
$f(T)$-gravity using different scenarios like, for example, fast extrasolar
planets orbiting  their parent stars at distances smaller than Mercury.
Furthermore, Earth-based  laboratory experiments (in which the simple
spherical geometry will not be the only case), where  the allowed
corrections are also expected to be small according to the recent tests
on General Relativity (Dimopoulos et al. \citep{Dimopoulos:2006nk},
Turyshev \citep{Dimopoulos:2006nkb}), would be worthwhile. However,
what could happen in even smaller distances is unknown, where the
possibility of large divergences from General Relativity could remain open.

\section*{Acknowledgments}
E.N.S wishes to thank K. Bamba, C. Bohmer, Y-F. Cai, S. Capozziello,  S.-H.
Chen,  J. B. Dent, S. Dutta R. Ferraro, F. Fiorini, P.A. Gonzalez, A.
Kehagias, M. Li,
J. W. Maluf, 	J. G. Pereira, M. E. Rodrigues, T. Sotiriou,   Y. Vasquez,
T. Wang and H. Wei
for useful discussions. The research project is implemented within the
framework of the Action «Supporting Postdoctoral Researchers» of the
Operational Program ``Education and Lifelong Learning'' (Action’s
Beneficiary: General Secretariat for Research and Technology), and is
co-financed by the European Social Fund (ESF) and the Greek State.

\end{document}